\documentstyle[]{mn}

\input{psfig.sty}

\ifnfsstwo

\fi

\ifnfssone
  \newmathalphabet{\mathit}
    \addtoversion{normal}{\mathit}{cmr}{m}{it}
    \addtoversion{bold}{\mathit}{cmr}{bx}{it}

\fi

\ifoldfss

\fi

\loadboldmathitalic
\loadboldgreek

\def\Zsun{\thinspace\hbox{$\hbox{Z}_{\odot}$}}
\def\msun{\thinspace\hbox{$\hbox{M}_{\odot}$}}

\def\kms{\,km~s$^{-1}$}      
\def\gal{galaxy}
\def\gals{galaxies}
\def\el{elliptical}
\def\sph{spheroid}

\def\ev{evolution}

\def\chdyn{chemodynamical}
\def\for{formation}
\def\sfor{star formation}
\def\gfor{galaxy formation}
\def\cf{cooling flow}
\def\gw{galactic wind}

\def\lbgs{Lyman break galaxies}
\def\C2{DSF 2237+116 C2}

\def\cgrd{colour gradient}
\def\mgrd{metallicity gradient}
\def\vi{$V_{606}-I_{814}$\/ }
\def\bv{$B_{450}-V_{606}$\/ }
\def\ub{$U_{300}-B_{450}$\/ }

\def\omm{$\Omega_m$}
\def\oml{$\Omega_{\Lambda}$}

\title[Colour gradients in spheroids]
       { The cosmological evolution of colour gradients in spheroids}
\author[A. C. S. Fria\c ca and R. J. Terlevich]
       {A. C. S. Fria\c ca$^1$ and R. J.
Terlevich$^{2,3}$\\
$^1$Instituto Astron\^omico e Geof\'\i sico, USP,
Av. Miguel Stefano 4200, 04301-904 S\~ao Paulo, SP, Brazil\\
$^2$Institute of Astronomy,
Madingley Road, Cambridge CB3 0EZ, UK\\
$^3$Visiting Professor at Instituto Nacional de Astrof\'{\i}sica,
Optica y Electr\'{o}nica. Av. Luis Enrique Erro 1, Tonanzintla, Puebla,
Mexico}

\pubyear{2000}
\begin{document}

\maketitle

\begin{abstract}
The analysis of the four-colour maps of galaxies in the Hubble Deep Field
has revealed in the sample of 0.4$<z<$1 early-type field galaxies
the existence of ellipticals with a  predominantly old coeval stellar population.
However, there is another, unexpected,
category of HDF early-type \gals, in which the \gal\ core
is significantly bluer than the outer regions.
We demonstrate that these colour gradients are predicted by 
the multi-zone \chdyn\ model for the \ev\ of \el\ \gals.

We suggest that the \cgrd\ could be used as a chronometer 
for the \ev\ of \el\ \gals: \gals\ younger than a few Gyr
exhibit  cores bluer than the surrounding \gal, due to ongoing \sfor\, while
more evolved \gals\ have redder cores, due to \mgrd s increasing
toward the centre.

\end{abstract}


\section{Introduction}

A surprising result of 
the analysis of the four-colour
($U_{300}B_{450}V_{606}I_{814}$) maps of the Hubble Deep Field
(Abraham et al. 1999, hereafter A99;
Menanteau, Abraham \& Ellis 2000)
is the existence of early-type field \gals, 
with positive \cgrd s in the centre
(\gal\ nucleus bluer than the outer surrounding region)
evidencing recent nuclear star formation 
in early-type galaxies with 0.4$<z<$1.

Although these bluer core \sph s (BCSs) 
could be explained within the scenario of
structure \for\ via hierarchical merging, in which 
the recent nuclear star formation would be the result of mergers
leading to the \for\ of an \el\ \gal, the high degree of symmetry 
in the images requires, additionally, that, after the merger, the systems had had
time to relax to a spheroidal geometry.

On the other hand, the appearance of these systems is exactly that
predicted by the \chdyn\ model for \ev\ of \el\ galaxies
(Fria\c ca \& Terlevich 1994; Fria\c ca \& Terlevich 1998, hereafter FT98).
The \chdyn\ model is a single collapse model, differing, however,
from the monolithic collapse model,
in that it is a multi-zone model, in which there is no coordination
in the \sfor\ in the several zones of the model.
This allows that, locally, the time scales for \sfor\ could be
of the order of 0.1 Gyr in the central regions of the \gal,
leading to a super-solar [Mg/Fe] ratio as observed in luminous \el s,
while it takes $\sim 1$ Gyr for the main body of the \el\ \gal\ to be formed.
An important ingredient of the \chdyn\ model is the persistence of
a central \cf\ for $2-3$ Gyr, feeding \sfor\ in the core,
which would account for the core being bluer than the rest of the \gal,
as in the BCSs found in the HDF.

Within the frame of the \chdyn\ model, the BCSs at intermediate redshifts
would be one of the manifestations of the general \ev\ of \sph s.
At higher redshifts, the chemodynamical model for the \ev\ of \sph s
predicts a link of the BCSs with Lyman break galaxies (Steidel 1995, 1996)
and the scarcity of passively evolving elliptical galaxies
(Zepf et al. 1997, Barger et al. 1999).
At $z \sim 3$, the first $\sim 1$ Gyr of our model \gals\ shows
striking similarities to the \lbgs\ (LBGs):
intense star formation, compact morphology, 
the presence of  outflows, and significant metal content.
Our investigation of the nature of the LBGs using
the chemodynamical model for \sph s (
Fria\c ca \& Terlevich 1999, hereafter FT99) supports a scenario in which
LBGs are the progenitors of  the present-day (age of 13 Gyr) 
moderately bright ($0.05 L^* - 1.4 L^*$) spheroids.
Over the $z > 1$ range in general,
the continuing (for $\sim 2-3$ Gyr) \sfor\ in the centre of \el s
predicted by the \chdyn\ model could explain
the relative absence of very red \gals\ in 
deep optical and near-infrared surveys, as expected from a population of 
passively evolving elliptical galaxies (Jimenez et al. 1999).
The BCSs would indicate the persistence of the same 
continuing central \sfor\ at redshift $<1$.
It should be noted, however, that most of the mass of the stellar
population is formed during the first Gyr; typically half of the
present stellar mass of the \gal\ is formed in $\sim 0.5$ Gyr,
and $\sim 5\%-30\%$ during the extended lower level
central \sfor\ phase. 

In this paper, we follow in detail the \ev\ of \cgrd s in \sph s
within the scenario of the \chdyn\ model and investigate the
relation of the continuing central \sfor\ activity to the BCS population.
Section 2 gives a description of the \chdyn\ model.
Section 3 presents the predicted \ev\ of \cgrd s in \sph s.
We show that, for a wide range of masses, there are 3 phases:
1) an earlier phase, with relatively flat \cgrd s;
2) the bluer core phase; and
3) the redder core phase.
Section 3 also discuss the possibility of using the transition
from the bluer core phase to the redder core phase as 
an evolutionary clock in cosmological studies.
Finally, in section 4, we identify the BCS seen at $0.4 < z < 0.7$
as sub-$L^*$ \sph s formed at $z>1$ with continuing central
\sfor\ or recurrent bursts of central \sfor.
Section 4 summarises our conclusions.

We adopt in this paper $H_0=70$ \kms\ Mpc$^{-1}$, in accordance with
the value obtained by HST Key Project on the Extragalactic Distance Scale,
$H_0=71 \pm 6$ \kms\ Mpc$^{-1}$, 
or, including a possible metallicity dependence of
the Chepheid period-luminosity relation, $H_0=68 \pm 6$ \kms\ Mpc$^{-1}$
(Mould et al. 2000).
We also adopt a flat $\Omega_m+\Omega_{\Lambda}=1$ cosmology.

\section{The chemodynamical model}

In this work, the \ev\ of \sph s is modelled with the
aid of the \chdyn\ model of FT98.
The chemodynamical model
combines multi-zone chemical evolution with 1-D hydrodynamics
to follow in detail the evolution and radial behaviour 
of gas and stars during the formation of a \sph.
Within each spherical zone, 
in which the model galaxy has been divided for the 1-D hydrodynamical calculations,
the evolution of the abundances of six chemical species (He, C, N, O, Mg, Fe) 
is calculated by solving the basic equations of chemical evolution.
The star formation and the subsequent stellar feedback regulate
episodes of wind, outflow, and cooling flow.
The knowledge of the radial gas flows in the galaxy allows us to
trace metallicity gradients,
and, in particular, the formation of a high-metallicity core in \el s.

FT98 built a sequence of  chemodynamical models reproducing the main
properties
of luminous elliptical galaxies.
The calculations begin with a gaseous protogalaxy
with initial baryonic mass $M_G$.
FT98 investigates models with $M_G$ between $10^{11}$ 
and $5\times10^{12}$ $\msun$.
In order to study LBGs, FT99 
have subsequently extended the mass grid of the models of FT98
downwards to $5\times 10^9$ \msun.
Intense star formation during the early stages of the \gal\  builds up
the stellar body of the \gal,
and during the \ev\ of the \gal, gas and stars exchange mass
through star formation and stellar gas return.
Owing to inflow and \gw\ episodes occurring during the \gal\ \ev,
its present  stellar mass is $\sim15-70$\% higher than $M_G$.
Gas and stars are embedded in a dark halo of core radius $r_h$
and mass $M_h$ (we set $M_h=3M_G$).
The models are characterised by $M_G$, $r_h$, and a \sfor\ prescription.
The SFR is given by a Schmidt law $\nu_{SF} \propto \rho^{x_{SF}}$ 
($\rho$ is the gas density and 
$\nu_{SF} = SFR/\rho$ is the specific SFR).
Here we consider the standard \sfor\ prescription of FT98, 
in which the normalization of $\nu$ is $\nu_0=10$ Gyr$^{-1}$
(in order to reproduce the super-solar [Mg/Fe] ratio of giant \el s)
and $x_{SF}=1/2$ (with the exception of one model, 
for which $x_{SF}=1/3$).
The stars form in a Salpeter IMF from 0.1 to 100 \msun.
A more detailed account of the models can be found in FT98.

 \begin{figure}
 \centerline{
 \psfig{figure=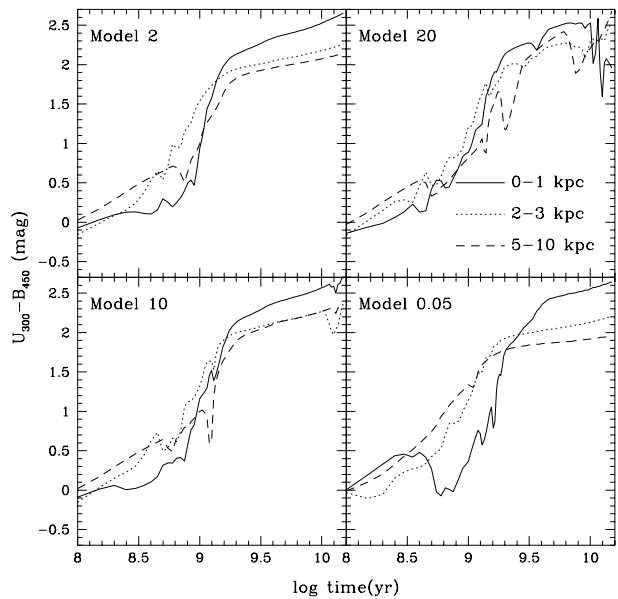,width=8.5cm,angle=0}}
 \caption{
Evolution of the colour \ub (rest-frame) through several annular apertures
for the models with $M_G=2\times 10^{11}$ \msun\ (labeled Model 2),
with $M_G=2\times 10^{12}$ \msun\ (Model 20),
with $M_G=10^{12}$ \msun\ (Model 10),
and with $M_G=5\times 10^{9}$ \msun\ (Model 0.05).
}
\end{figure}

\section{Evolution of colours gradients}

We have used the spectrophotometric models
of Bruzual \& Charlot (1998) to compute the SED
for each \gal\ zone. This is done in a self-consistent way 
using the metallicity given for each zone and each time by the
\chdyn\ model and for a consistent IMF.
Then, the SED is integrated  over several apertures,
redshifted,
reddened by the Lyman $\alpha$ forest opacity (Madau 1995), 
and convoluted with the filter transmission curves. 
Unless otherwise stated, AB magnitudes are used throughout this paper.

It is helpful for the understanding the \ev\ of the \cgrd s
to consider first the \ev\ of the colours
in the rest-frame of the \gal.
Figure 1 shows the \ev\ of the \ub colour,
more sensitive to on-going \sfor,
as seen through several annular apertures,
for several models, with $M_G$ ranging from $5\times 10^{9}$
to $M_G=2\times 10^{12}$ \msun.
The \ev\ of model with $M_G=2\times 10^{11}$
(the fiducial model of FT98, or model 2
---following FT98 we label the models according $M_G$
in units of $10^{11}$ \msun)
is typical of models covering a wide range of 
masses (from less than $10^{10}$ up to $10^{12}$ \msun). 
Its \ev\ exhibits 3 distinct epochs regarding the \cgrd s:
1) An early phase, during which the \gal\ as a whole is blue,
due to wide spread intense  \sfor. The region with the highest SFR's
extends out to $r\sim 3$ kpc with little radial gradient of specific SFR,
which results in a \cgrd\ flat in this region.
The very central region can be even redder than 
its immediate surroundings during this phase,
due to the very fast chemical enrichment for $r<1$ kpc
(where, the metallicity, at 0.1 Gyr, of the stellar population is 0.6 \Zsun)
2) After $3\times 10^8$ yr, the SFR decreases from outside in,
but it persists for more than 1 Gyr in the core ($r<1$ kpc),
which is significantly bluer than the rest of the \gal:
that is the {\it bluer core} phase, during which
the \gal\ would be seen as a BCS;
3) The last phase ($t > 1$ Gyr) is the {\it redder core} phase,
during which the level of \sfor\ is  low in the core,
which is redder than the outer parts of the \gal,
due to its high metallicity.
The dips in the \ev\ of \ub indicates bursts of \sfor.
Note that the transition from bluer core phase
to the redder core phase is rapid, which suggest
the use of this transition as a time-marker.
The transition occurs later if we consider the surrounding
$2<r<3$ kpc region, because in the outermost radii ($5<r<10$ kpc),
the lower metallicities imply bluer colours.

The \ev\ of the model with 
$M_G=2\times 10^{12}$ \msun\ (model 20 of FT98)
illustrates another sequence of
phases that occurs only for very massive systems
($M_G \ga 10^{12}$ \msun). 
In this mass range, the \gw\ occurs late, and the
inner \cf\ stops for a short time after which
it reappears and eventually becomes a massive,
global \cf\ extending over 100 kpc.
The bluer core is clearly present only in the
early stages of the \ev\ (till $\sim 0.3$ Gyr).
After this time, due to the complex inner \cf\ behaviour,
which extends over the whole $r<10$ kpc,
several uncoordinated \sfor\ episodes rend the
\cgrd s ill-defined. 
For $t \ga 1.5$ Gyr, the core is redder than
its surroundings, although the transition of 
a bluer core to a redder core is not smooth,
and the \ev\ of the colour of the core is
subject to some blueing at $\sim 3$ Gyr
due to a weak, late, central \sfor\ episode.
Finally, in late times $t \la 10$ Gyr, the 
establishment of a massive, global \cf\ recovers
a bluer core.

Note that, although the model with $M_G=10^{12}$ \msun\ is also
massive enough to develop a \cf\ in $t>10$ Gyr,
(visible as a dip in the \ub profiles after this time),
during its first 10 Gyr it exhibits the three-phase \ev\ of
the \cgrd s, with the characteristic clear separation 
between the bluer core and the redder core stage at $\sim 1$ Gyr.

 \begin{figure}
 \centerline{
 \psfig{figure=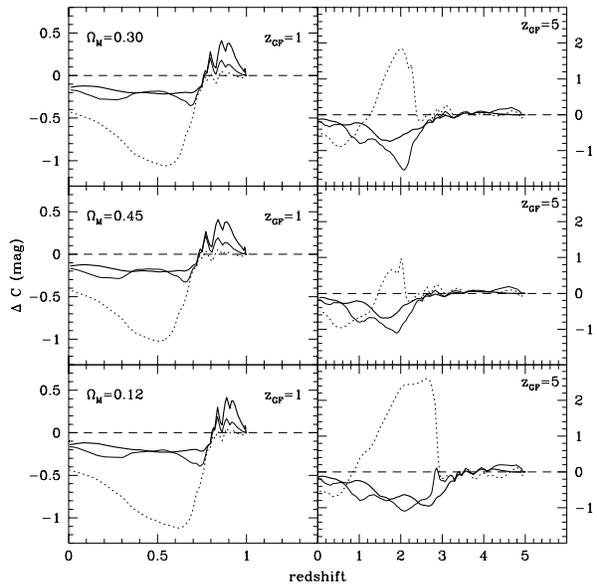,width=8.5cm,angle=0}}
 \caption{
The \ev\ of the \cgrd s $\Delta C$ of the colours:
\vi (heavy solid lines), \bv (light solid lines), 
and \ub (dotted lines),
for the model with $2\times 10^{11}$ \msun and
for several cosmologies and two epochs ($z_{GF}$) of \gfor.
}
\end{figure}

\subsection{A typical $\sim L^*$ galaxy}

The fiducial model of FT98
($M_G=2\times 10^{11}$ \msun\ and $r_h=3.5$ kpc )
has a present-day stellar mass, $2.4\times 10^{11}$ \msun,
corresponding to $L_B = 0.7 L^*$, being typical for
a present-day moderately luminous \el\ \gal.
Figure 2 shows the \ev\ of the \cgrd\ for this model.
The quantity $\Delta C$ is defined as the colour
at an annulus $5<r<10$ kpc minus the colour
in the central kpc ($r<1$ kpc). Positive values of $\Delta C$
therefore correspond to positive central \cgrd s,
and indicate a  core bluer than the surrounding \gal.
The $5<r<10$ kpc dimension chosen for the annulus representing
the non-nuclear part of the galaxy falls inside a typical
effective radius of a luminous elliptical 
($\sim 10$ kpc for a $L^*$ \el).
This region is easily resolvable 
by large ground-based telescopes,
since 1 arcsec corresponds to 8.0 (9.1, 7.4) kpc
for $H_0=70$ \kms\ Mpc$^{-1}$ and $\Omega_m=0.3$ ($=0.12$, $=0.45$)
(assuming a flat cosmology $\Omega_m+\Omega_{\Lambda}=1$).
At the same time, the $r<10$ kpc region avoids
the $\propto (1+z)^{-4}$ Tolman dimming of the surface brightness
which makes it difficult to detect 
the outermost regions of  the \gal, if it is at high redshifts.
As we can see from Figure 2, 
the changes in the central \cgrd\ are more dramatic for an epoch of
\gfor\ $z_{GF}=5$, but they are very noticeable for $z_{GF}=1$.
If the \gal\ is formed at a recent epoch ($z_{GF}=1$), the bluer
core is more noticeable in the \vi colour, while for higher redshifts
of \gfor\ , the \ub colour 
would trace better the continuing \sfor\ in the \gal\ centre.
For instance, for $z_{GF}=5$, the \gal\ would have an bluer core
for $z < 3$ (when it is in the phase of central continuing \sfor)
only in the \ub colour 
(with a maximum around $z=2$ for $z_{GF}=5$),
while for the \bv and \vi colours, the core is $redder$ than
the surrounding \gal.

The bluer core phase may be characterised by $z_{rev}$, 
the redshift at which
there is the reversal from blue core (due to central \sfor)
to red core (due to \mgrd).
The values of $z_{rev}$ are exhibited in Table 1
for the models, colours and cosmologies used in this paper.

The colour \vi (in which we can follow more closely the \ev\ of the \cgrd)
has $z_{rev}=0.764$ for a $\Omega_m=0.3$, $\Omega_{\Lambda}=0.7$ cosmology
and a recent \gfor\ epoch ($z_{GF}=1$).
The maximum of \vi is $\Delta C=0.414$ at $z=0.859$.
There is also a secondary maximum of the central \sfor\ at $z=0.799$
($\Delta C=0.285$). At the maximum of \sfor, this \gal\ would be very
bright ($I_{814}=17.74$) and it would be easily detectable in the HDF.
Since no BCS as bright as this has been detected in A99 sample,
we can tentatively conclude that the typical BCSs are not $\sim L^*$-\el s
formed at $z_{GF}\sim 1$.
Turning to higher $z_{GF}$'s ($z_{GF}=5$), the blue core is more
pronounced in the \ub, in which the maximum \cgrd\ occurs
at $z=2.012$ ($\Delta C=1.851$).
Note, however, that the model \gal\ would be practically invisible
in the $U_{300}$ band ($U_{300}=29.51$ for the whole \gal),
while still detectable in other bands 
($B_{450}=25.66$, $V_{606}=24.55$, and $I_{814}=23.71$).
As a matter of fact, it would be classified as a LBG,
because its colours \bv=1.11 and \ub=3.85 satisfy the
criteria for selecting LBGs at $2 < z < 3.5$:
\bv$\leq 1.2$ and \ub$>$\bv$+1.0$ (Dickinson 1998).
Note that for a $\Omega_m=0.12$, $\Omega_{\Lambda}=0.88$ cosmology,
the luminosity distance would be larger and the \gal\ 
fainter: $U_{300}=31.05$, $B_{450}=27.97$, $V_{606}=26.23$, 
and $I_{814}=24.99$ at $z=2$. 
Then, \bv=1.74 and it would no longer be identified as a LBG.

\subsection{A massive central galaxy in a cooling flow}

 \begin{figure}
 \centerline{
 \psfig{figure=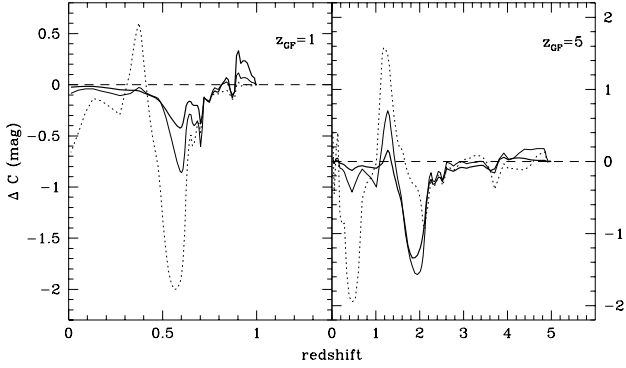,width=8.5cm,angle=0}}
 \caption{
The same as Figure 2, but
for the model with $2\times 10^{12}$ \msun.
A $\Omega_m=0.3$, $\Omega_{\Lambda}=0.7$ cosmology
is adopted here.
}
\end{figure}

Figure 3 shows the \ev\ of the \cgrd\ for 
the FT98 model with $M_G=2\times 10^{12}$ \msun,
which has a present-day stellar mass of $2.4\times 10^{12}$ \msun.
Note that the late \cf\ episode in this model would
require enough gas supply through the outer boundary of the \gal,
a condition that would be met if the \gal\ would be located at
the centre of a relaxed cluster of \gals.
As a matter of fact, the model with $M_G=2\times 10^{12}$ \msun\ could
represent an \el\ at the bottom of the potential well
of a cluster of \gals\ collecting
the gas of a \cf\ formed in the intra-cluster medium,
since the most massive central \gals\  of \cf s 
have masses of this order.

On the other hand, the early \cf\ episode could be maintained
by the gas within the dark mass halo of the protogalaxy itself
and thus would not require that the \gal\ be located in a cluster,
or, more properly speaking, that the protocluster 
containing the \gal\ be well-developed.

In this model, the first episode of \sfor\ is over a more extended region
so the central \cgrd\ is smaller.
In addition, for a \gal\ formed at $z_{GF}=5$, the reversal
blue core/red core is followed by a contra-reversal red core/blue core
due to later central \sfor, which would correspond to the present-day
central \sfor\ activity seen in a few central dominant \gals\ in \cf s
(McNamara 1997).

At late times, Model 20 shows starbursts in its core fed by the late
\cf. As a consequence, as we approach $z\sim 0$, $\Delta C$ varies
wildly, being positive most of the time (\sfor\ in the  core prevails).

\subsection{Recurrent star formation in ellipticals}

 \begin{figure}
 \centerline{
 \psfig{figure=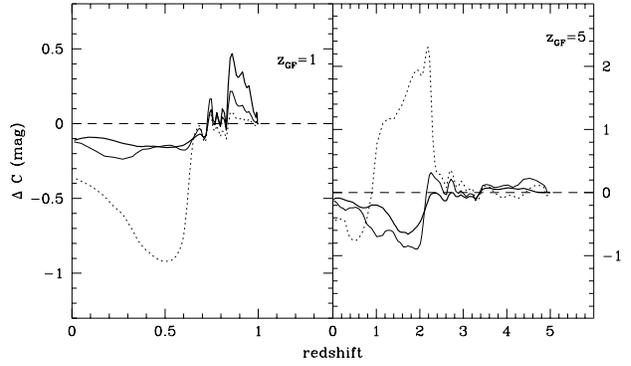,width=8.5cm,angle=0}}
 \caption{
The same as Figure 3, but
for the model with $2\times 10^{11}$ \msun\ and $x_{SF}=1/3$.
}
\end{figure}

One particularly interesting model is the model 2(1/3) of FT98
because it exhibits recurrence of late central inflow episodes.
This model has $M_G=2\times 10^{11}$ \msun, as the fiducial model,
but $x_{SF}=1/3$ instead of $x_{SF}=1/2$
--- $x_{SF}$ is the index of Schmidt law describing the dependence
of the specific star formation rate on the gas density 
$\nu_{SF}\propto \rho^{x_{SF}}$ (see end of Section 2).
As the SN I rate decreases, the wind stalls, the gas starts to accumulate
in the \gal\ core, and the radiative losses drive a new central \cf.
The late central inflow episodes are brief
and quickly put out by the resulting burst of star formation.
The duration of the late \cf\ episodes is a few times $10^7$ yr,
and, since the separation between them is typically a few Gyr,
their duty cycle is $\sim 1/100$.
The turning on and off of the central inflow could correspond to the
``active" and ``inactive" states of the core in
both the super-massive black hole model of AGN
and the starburst model of AGN.
In addition, the length of the duty cycle of the central inflow
increases with time and the amount of material deposited
decreases, implying a steep decrease of the time-averaged
luminosity of any central activity powered by the infalling material.
For the model 2(1/3),  the second central inflow occurs
at $t=4$ Gyr and involves $1.5\times 10^7$ $\msun$,
and the third one ($8.4\times 10^6$ $\msun$) happens at $t=9.5$ Gyr.
Each late central inflow episode is accompanied by enhanced \sfor\ inside
the central kpc, the signature of which is a temporary  blueing
of the \gal\ core.

These late episodes of \sfor\ of the model 2(1/3) imply that
the blue core persists even at redshifts lower than
in the fiducial model. This can be  seen from the lower $z_{rev}$ 
of the model 2(1/3) in comparison
to that of model 2.
For $z_{GF}=1$, the first central \cf\ episode is more conspicuous
in the \vi, as was the case for model 2.
The maximum of the \vi \cgrd\ occurs at $z=0.853$ when $\Delta C=0.473$
(in the \bv colour, $\Delta C=0.222$).
However, the second central \cf\ episode is more noticeable in \bv,
corresponding to the local maximum $\Delta C=0.179$ at $z=0.744$
($\Delta C=0.114$ in \vi).
The third inner \cf\ episode has a more pronounced signature in \ub,
but the associated starburst is so weak 
that it does not change the redder core to a bluer core,
so $\Delta C=-0.012$ at $z=0.681$.
The \mgrd\ effect dominates over the impact of the \sfor,
as we can see from nearly constancy of the 
\vi\ \cgrd\ ($\Delta C=-0.071$ at $z=0.681$).

\begin{table*}
\caption{The redshift of reversal from blue core to red core}
\begin{tabular}{llcccrcccrccc}
\hline
&&\multicolumn{3}{c}{Model 2}&&\multicolumn{3}{c}{Model 2(1/3)}
&&\multicolumn{3}{c}{Model 20}\\
$\Omega_m$&$z_{GF}$&$U-B$&$B-V$&$V-I$&&$U-B$&$B-V$&$V-I$&&$U-B$&$B-V$&$V-I$\\
\hline
0.12&1&0.835&0.805&0.809&&0.780&0.777&0.779&&0.388&0.852&0.851\\
0.30&1&0.792&0.759&0.764&&0.729&0.725&0.730&&0.311&0.814&0.814\\
0.45&1&0.763&0.729&0.735&&0.695&0.693&0.698&&0.268&0.789&0.791\\[3pt]

0.12&2&1.640&1.746&1.589&&0.927&1.444&1.536&&\dots&\dots&\dots\\
0.30&2&1.517&1.463&1.450&&1.176&1.272&1.382&&0.624&0.691&1.575\\
0.45&2&1.447&1.366&1.370&&1.089&1.190&1.296&&0.528&0.598&1.506\\[3pt]

0.12&3&1.093&2.367&2.372&&0.882&1.958&2.023&&\dots&\dots&\dots\\
0.30&3&1.690&2.325&2.001&&0.913&2.305&1.898&&\dots&\dots&\dots\\
0.45&3&1.886&2.227&1.865&&1.378&2.198&1.755&&0.678&0.758&2.452\\[3pt]

0.12&5&0.853&2.823&3.540&&0.772&2.416&2.789&&\dots&\dots&\dots\\
0.30&5&1.263&2.864&3.438&&0.899&2.153&2.713&&\dots&\dots&\dots\\
0.45&5&1.457&2.752&3.240&&0.868&1.959&2.059&&\dots&\dots&\dots\\
\hline
\end{tabular}
\end{table*}

\subsection{Colour gradients as cosmological probes}

Star formation activity in high redshifts \gals\ has already been
proposed to be used to determine the cosmological constant (\oml)
and the matter mass density of the Universes (\omm)
(Melnick, Terlevich \& Terlevich 2000).
We suggest here the use of \cgrd s as a cosmological probe.
The time-scales for the \ev\ of the central \cgrd\ in \sph s
can potentially provide an independent clock for measuring the age of the Universe.
Figures 2 also exhibits the results for three different cosmologies,
illustrating the sensitivity of the \ev\ of the \cgrd\ to
the value of \omm. 
We consider a flat cosmology $\Omega_m+\Omega_{\Lambda}=1$
with $0.12 \leq \Omega_m \leq 0.45$,
as favoured by the simultaneous constraints of 
the first acoustic peak of the angular power spectrum
of the cosmic radiation background measured by the BOOMERang
(de Bernardis et al. 2000) 
and the redshift-magnitude relation derived from high redshift supernovae
surveys (Perlmutter et al. 1999, Riess et al. 1998, Schmidt et al. 1998).
These latter works have already found evidence for an accelerating Universe
in which matter would provide only $\sim 30$ \% of the critical density
for a flat Universe, with the rest being accounted for by the cosmological
constant $\Lambda$.
Therefore, the three cosmologies adopted here consider
$\Omega_m=0.3$, $=0.12$ and $=0.45$.

An additional constraint is the age of the Universe given
by each cosmology: 11.98, 13.47 and 17.09 Gyr,
for $\Omega_m=0.45$, $=0.3$ and $=0.12$, respectively.
The $\approx 10\%$ errors on $H_0$ prevent us from 
using values of $H_0$ significantly lower than $70$ \kms\ Mpc$^{-1}$,
so the age of the Universe implied by $\Omega_m=0.45$
would be dangerously low in comparison to the ages
of the oldest globular clusters, $\sim 11.5-14$ Gyr,
derived in the post-HIPPARCOS age
(D'Antona, Caloi and Mazzitelli, 1997; 
Salaris, degl'Innocenti and Weiss, 1997; 
Pont et al. 1998; Chaboyer et al. 1998; Carretta et al. 2000).

We should also take into account the constraints on the age of the Universe
from high redshift old \gals\ (Lima \& Alcaniz 2000).
One of the strongest lower limits on the age of the high redshift Universe
comes from the very red \gal\ 53W069 at z=1.49, with a minimal
stellar age of 4.0 Gyr (Dunlop 1999), which, for $H_0 > 60$ \kms\ Mpc$^{-1}$,
implies $\Omega_m<0.5$.

A useful quantity that could be used to set the colour clock
of the evolving \sph\ and distinguish between 
several cosmologies and investigate the epoch of \gfor\ is $z_{rev}$.
The cosmologies with smaller $\Omega_m$'s (larger $\Omega_{\Lambda}$'s)
have larger lookback times, and therefore, higher $z_{rev}$'s.

Moreover, the models with masses around that of an $L^*$-\el\ progenitor
($M_G=10^{11}-10^{12}$ \msun) have $z_{rev}$ remarkably similar,
which guarantees a fair accuracy for the \cgrd\ clock.
For a $\Omega_m=0.3$, $\Omega_{\Lambda}=0.7$ cosmology,
the model with $10^{11}$ \msun\ has 
$z_{rev}=0.745$ ($z_{GF}=1$ and \vi) 
and $z_{rev}=1.159$ ($z_{GF}=5$ and \ub).
The respective values for the model with $10^{12}$ \msun\ are 0.806 and 0.970,
and, for model 2, 0.764 and 1.263.

Model 20 is not particularly useful as a colour clock,
due to the massive late \cf, which prevails at the present time.
It is worth to note that if the model is old enough
($z_{GF}$ high and $\Omega_m$ low), there is central \sfor\ driven
by the late \cf.
Therefore, as we can see from Table 1, the model \gal\ has
a bluer core at the present day for cosmologies and $z_{GF}$'s
that imply larger lookback times till the epoch of \gfor, $t_{b,GF}$, 
and in this case $z_{rev}$ is not defined.

\subsection{Bluer Core Spheroids}

 \begin{figure}
 \centerline{
 \psfig{figure=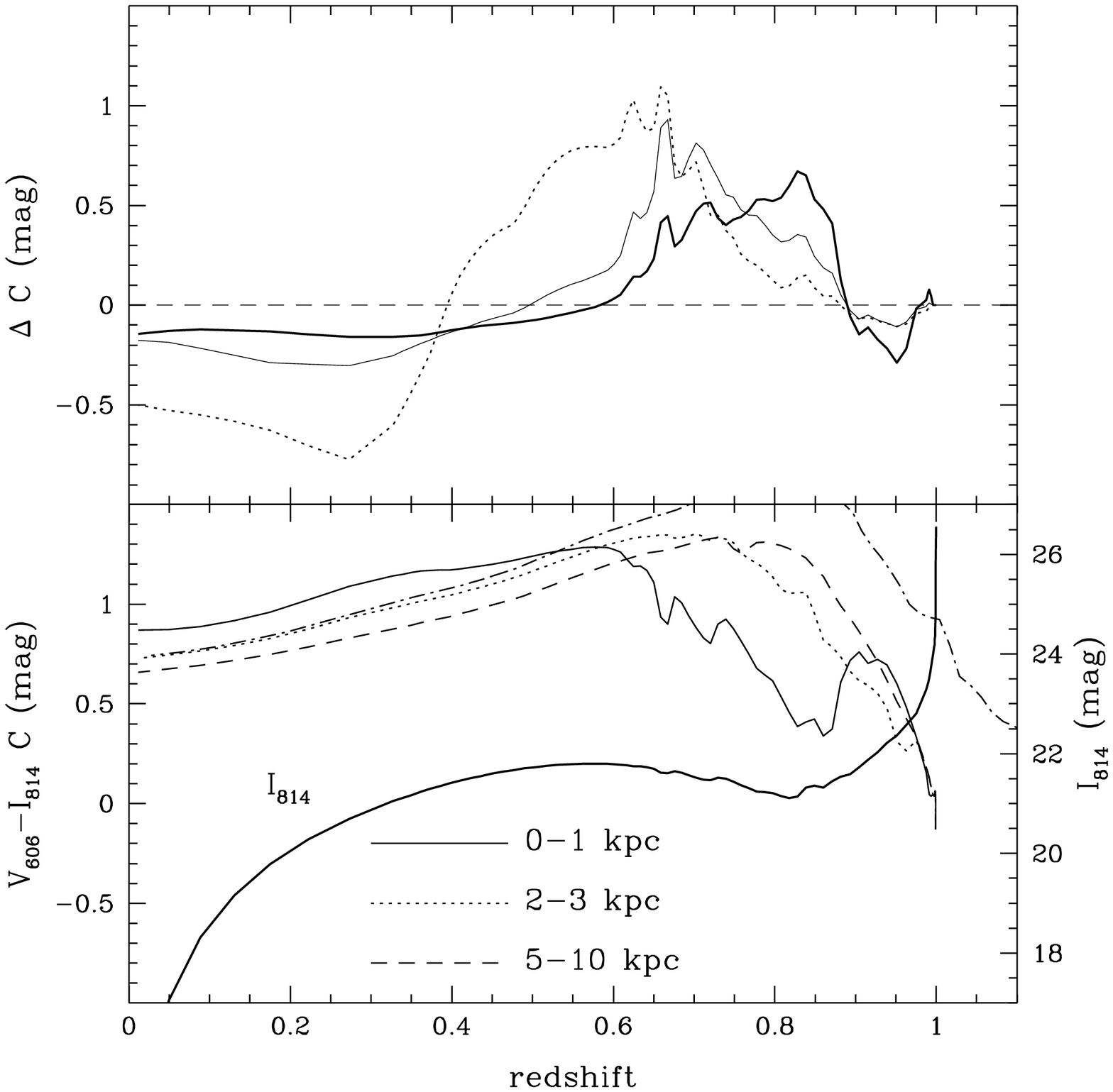,width=8.5cm,angle=0}}
 \caption{
Top panel: \ev\ of the \cgrd s $\Delta C$ of the colours:
\vi (heavy solid lines), \bv (light solid lines), 
and \ub (dotted lines),
for the model with $M_G=5\times 10^{9}$ \msun, 
for the $\Omega_m=0.3$, $\Omega_{\Lambda}=0.7$ cosmology 
and an epoch of \gfor\ $z_{GF}=1$.
Here $\Delta C$ is defined as the colour
seen inside the annulus $2<r<3$ kpc minus the colour
in the central kpc ($r<1$ kpc).
Lower panel:
evolution of the \vi colour through several annular apertures.
The dot-dashed line represents the \vi colour through
a $2<r<3$ kpc aperture for $z_{GF}=1.2$.
Also given the \ev\ of $I_{814}$ for the whole \gal.
}
\end{figure}

It is interesting that the lower mass models of FT99,
intended to describe typical LBGs with velocity dispersions around
70 \kms\ have properties that are similar to those
of typical BCSs of A99 (for instance, BCS HDF-2.264 at z=0.478).
In particular, the models with $M_G=$ $10^{10}$ \msun\ and 
$5\times 10^9$ \msun\ have prolonged central \sfor\ extending
for more than 3 Gyr (see Figure 1 of FT99).
Because of this, they could be seen as BCSs over a significant
fraction of their lifetime.
These objects, if they have been formed at $z_{GF}\sim 4-5$,
would be LBGs at $z\sim 3$, but if $z_{GF}\sim 1$ instead,
they would be classified as BCSs at $z \sim 0.7$.

Figure 5 illustrates the \ev\ of the \cgrd s of the FT99 model
with $5\times 10^9$ \msun\ (hereafter model 0.05), assuming $z_{GF}=1$.
Model 0.05 would be seen as a $\sigma =55$ \kms\ LBG at an age of 0.5 Gyr,
and as a present-day (age of 13 Gyr) $0.05 L^*$ \sph.
In this model, the central \cf\ is
extinguished only at 3.63 Gyr. By comparison, in the fiducial model 2,
the inner \cf\ ends at 1.80 Gyr.
In order to compare our results with observations with
the same resolution as those of HST as in A99, 
now $\Delta C$ is defined as the colour
inside the  annulus $2<r<3$ kpc minus the colour in the central kpc,
since the WFPC-2 pixel size of $0.1^{\prime\prime}$ corresponds 
to $0.6$ kpc at $z=0.5$ for a $\Omega_m=0.3$, $\Omega_{\Lambda}=0.7$ cosmology.
The resolution is even better in the colours $V-I$ maps of A99,
which have a  drizzled pixel scale of $0.04^{\prime\prime}$,
allowing, for instance, a clear definition of the 
$r\approx 0.25^{\prime\prime}$ blue core of the BCS HDF-2.264.
In addition, in this section, we use Vega magnitudes, as in A99.

As we see from Figure 5, for $z>0.6$, our model \gal\ would be easily detected
as a BCS in the \vi colour, which is the more reliably determined colour in A99
(Menanteau et al. 2000).
At $z=0.7$, $\Delta C$=0.472 and $I_{814}=21.52$,
while the \cgrd\ is flat in the outer part 
of the \gal\ (\vi(2-3 kpc)=1.354 and \vi(5-10 kpc)=1.311).
Note that the \gal\ is detectable as a BCS at lower redshifts
in \bv and \ub ($z_{rev}=0.583$, 0.496 and 0.395 in
\vi, \bv and \ub, respectively).
Model 0.05 was not tailored to reproduce the BCS HDF-2.264,
but the similarities of the properties of the model to the object
suggest that, if HDF-2.264 is typical of BCSs, they would-be
$z\sim 0.5$ counterparts of sub-$L^*$ \sph s, since
$I_{814}=21.73$ of the model at $z=0.5$ is comparable to 
$I_{814}=21.66$ (Menanteau et al. 2000) --- $I_{814}$
of our model varies within the range $21.2-21.8$ between $z=0.4$ and $z=0.8$.

The colours of the outer region ($2-3$ kpc) constrain
the average age of the stellar population of the \gal.
Within the scenario of our model
the colours of the outer parts of HDF-2.264
suggest $z_{GF}$ somewhat higher than 1.0.
As a matter of fact, model 0.05 has 
\vi(2-3 kpc)=1.171 at $z=0.5$ for $z_{GF}=1$,
while it has \vi(2-3 kpc)=1.223 at $z=0.5$ for $z_{GF}=1.2$,
thus reproducing well the red outskirts of HDF-2.264,
for which the reddest pixels have \vi$\approx 1.3$,
with an error of 0.07 mag (see Figure 4 of A99).

On the other hand, the recent central \sfor\  at $z=0.478$
of HDF-2.264 could be explained by some secondary central
infall similar to that of model 2(1/3).
Note that, at $t=1$ Gyr, the model 0.05 has a fully developed \gw\ with
1200 \kms\ at $r=10$ kpc. However, at $=3$ Gyr, the \gw\ has weakened 
(wind velocity of 270 \kms\ at $r=10$ kpc) and, in the inner regions
there is only a highly subsonic outflow (20 \kms\ at $r=1$ kpc).
The gas accumulated in the dark halo of the
\gal\ brakes the outflowing gas, so it is possible that, 
in some cases,
the core accumulates enough gas returned by 
the evolving stars to trigger late central \sfor\ episodes.
(this happens in model 2(1/3) but for a $\sim L^*$ \gal).
In this connection, we could explain the scarcity of BCSs
in \gal\ cluster environments (Menanteau et al. 2000)
as due to the stripping of the
gas in the dark halo of the \sph, thus preventing
very prolonged ($>3$ Gyr) \cf-fed central \sfor\ as well as
late recurrent central \sfor\ episodes.

\section{Conclusions}

The early-type galaxies with blue cores of A99 are very symmetrical,
showing no morphological peculiarity.
A99 points out that the $\sim 40\%$ fraction they find for 
the early-type systems
with recent \sfor\ is consistent with the predictions from hierarchical models.
However, in the hierarchical model scenario, after a merger event,
the resulting system has to relax quickly enough in order to be classified
as an early-type galaxy.

A alternative scenario to the merger picture would be
that proposed by the multi-zone single-collapse model,
in which late gas central inflow would feed \sfor\ in the
core of the \gal. 
A spherically symmetrical star forming population
would naturally develop in the centre of the \gal, 
and one would not need the extra ingredient
of a short relaxing time-scale as in the merger scenario.

The central blue cores arise naturally in our models,
in this way reproducing the \cgrd s which are typical of BCSs.
Within the scenario of our model,
if the HDF-2.264 BCS at z=0.478 is typical of BCSs, then they are
$z\sim 0.5$ counterparts of sub-$L^*$ \sph s,
formed at redshifts somewhat larger than $z_{GF}=1$,
and requiring very prolonged ($>3$ Gyr) \cf-fed central \sfor\ or 
recurrent central \sfor\ episodes.
This could account for the scarcity of BCSs
in rich clusters due to the removal of the
gas in the dark halo of the \sph\ by the intracluster medium.
Without the gas reservoir in the dark halo,
the gas inside the \gal\ would escape very rapidly,
thus preventing any long term \cf\ that could feed 
very prolonged central \sfor\ activity.
At the same time, 
any late central starburst would be suppressed, since
it could not be feed either by gas
ejected by evolved stars and accumulated in the \gal\ or
by old \gw\ ejecta braked in the gas-filled dark matter halo
and falling back towards the \gal.

If the progenitors of $\sim L^*$ \sph s were formed 
at redshifts larger than $z_{GF}=1$, they would
give rise to high-mass, high-redshift counterparts of the
$0.4 < z < 1$ BCSs.
If they are formed at $z_{GF}\sim 5$, they could be missed
in the standard multi-colour surveys designed to
select candidates to LBGs. 
Although it is possible to find some BCSs in the present LBG samples,
the colour criteria used in searching LBGs
should be made more flexible (for instance, redder \bv colour indices)
in order to search for 
high-redshift BCSs which would be the $\sim 0.3-$a to  few Gyr
old progenitors of the present day $\sim L^*$ \el s.
Moreover, according to the \chdyn\ model for \ev\ of \sph s,
due to the fact that the 
prolonged central \sfor\ characteristic of BCSs,
extends at least 2 Gyr, and, sometimes, longer than 3 Gyr,
the BCS stage in the \ev\ of \sph s lasts somewhat longer
than the LBG stage ($ 0.2 \la t \la 1.5$ Gyr) (FT99)
and should also be found in abundant numbers at high redshifts,
provided that we explore regions in the colour diagrams wider than those
used in the selection of LBG candidates, and perform deeper imaging
with higher resolution of the candidates to BCSs.
Finally, these high-redshift BCSs would be useful probes
for the epoch of \gfor\ and the cosmological parameters of the Universe.

\section*{Acknowledgments}

We thank Gustavo Bruzual for making us available the GISSEL code
for evolutionary stellar population synthesis.
We are grateful to the referee for a number of suggestions
that made this paper more readable.
A.C.S.F. acknowledges support from the Brazilian agencies
FAPESP, CNPq, and FINEP/PRONEX.

\bsp

\end{document}